\documentclass[final,5p,times,twocolumn]{elsarticle}
\usepackage{amsmath}

\usepackage{graphicx}

\usepackage{amssymb}
\usepackage{amsthm}

\biboptions{comma,square,sort&compress}

\newcounter{bla}

\usepackage[hidelinks,pdfauthor=author]{hyperref}
\usepackage{mathtools}
\usepackage{soul}
\usepackage{bm}
\usepackage{subfigure}
\usepackage{tikz}
\usetikzlibrary{trees}
\usepackage{enumitem}
\usepackage{booktabs}
\usepackage{url}

\journal{Computer Physics Communications}

\begin{document}
\begin{frontmatter}

\title{PARPHOM: PARallel PHOnon calculator for Moir\'{e} systems}

\author[a]{Shinjan Mandal}
\author[b]{Indrajit Maity\corref{IM_pres}}
\author[a]{H. R. Krishnamurthy}
\author[a]{Manish Jain\corref{author}}
\cortext[IM_pres]{Current Address: Max-Planck Institute for the Structure and Dynamics of Matter, Hamburg}
\cortext[author]{Corresponding author,
                  \textbf{E-mail address:} mjain@iisc.ac.in}
\address[a]{Centre for Condensed Matter Theory, Department of Physics, Indian Institute of Science, Bangalore}
\address[b]{Departments of Materials and the Thomas Young Centre for Theory and Simulation of Materials, Imperial College London}

\begin{abstract}

The introduction of a twist between two layers of two-dimensional materials has opened up a new and exciting field of research known as twistronics.
In these systems, the phonon dispersions show significant renormalization and enhanced electron-phonon interactions as a function of the twist angle.
However, the large system size of the resulting moir\'{e} patterns in these systems makes phonon calculations computationally challenging. 
In this paper, we present PARPHOM, a powerful code package designed to address these challenges. 
PARPHOM enables the generation of force constants, computation of phononic band structures, and determination of density of states in twisted 2D material systems. 
Moreover, PARPHOM provides essential routines to investigate the finite temperature dynamics in these systems and analyze the chirality of the phonon bands. 
This paper serves as an introduction to PARPHOM, highlighting its capabilities and demonstrating its utility in unraveling the intricate phononic properties of twisted 2D materials.
\end{abstract}


\end{frontmatter}

{\bf PROGRAM SUMMARY}

\begin{small}
\noindent
{\em Program Title:} PARPHOM \\
{\em Developer's repository link:} \url{https://github.com/qtm-iisc/PARPHOM} \\ 
{\em Licensing provisions:} GNU General Public License v3.0 (GPLv3)  \\
{\em Programming language:} FORTRAN, Python        \\
{\em External Routines/libraries:} numpy, LAMMPS (serial python wrapper), mpi4py, ScaLAPACK, HDF5, matplotlib, scipy, Spglib\\
{\em Nature of problem:} Due to the large number of atoms in 2D moir\'e systems, performing phonon calculations is quite challenging. The exorbitantly high memory requirements of such calculations make them infeasible with currently available codes.\\
{\em Solution method:} A parallel algorithm to generate the force constant matrices for these large moir\'e systems has been implemented. Parallel diagonalization routines available in ScaLAPACK are then used to diagonalize the dynamical matrices  constructed from the force constants at each $\mathbf{q}$ points. 
 \end{small}

\section{Introduction}
\label{Introduction}
Phonons are quantized lattice vibrations that play a crucial role in determining the structural, thermal, and transport properties of materials. Non-invasive spectroscopic techniques, such as infrared and Raman spectroscopy, enable direct probing of phonons in materials, providing insights into the impact of strain, doping, thickness, temperature, pressure, and defects ~\cite{Malardraman2009, Saitoraman2016}. First-principles calculations based on density functional theory (DFT) are widely used to compute phonon frequencies in crystals ~\cite{alfe2009phon,phonopy}. These calculations have been shown to accurately reproduce phonon frequencies in many materials \cite{baroni2001phonons}. However, despite advances in high-performance computing, DFT-based phonon calculations are limited to crystals with unit cells containing a few hundred atoms.

The twisting of two-dimensional (2D) materials relative to each other provides a unique degree of freedom for tuning their electronic \cite{ cao2018unconventional, cao2018correlated, lisi2021observation, ribeiro2018twistable, PhysRevB.95.075420, PhysRevLett.124.076801}, excitonic~\cite{tran2019evidence,wang2019evidence}, and phononic properties~\cite{carozo2011raman,zhao2013interlayer,lin2018moire,jiang2012acoustic,wang2013resonance}. 
At small twist angles, significant atomic relaxations lead to localized phonon modes and a strong renormalization of phonon frequencies \cite{quan2021phonon,ni2019soliton,debnath2020evolution,gadelha2021localization}.
Experimentally relevant moir\'{e} unit cells can contain hundreds to thousands of atoms, making standard first-principles approaches for computing phonons with relaxation effects prohibitively expensive. Recently, density functional theory (DFT)-fitted classical force-field-based calculations of phonons have been shown to accurately capture the renormalization of phonon modes in these systems\cite{lamparski2020soliton,ultrasoft_phason}. This approach provides a computationally manageable method for computing phonons in systems containing thousands of atoms.

Motivated by the great interest in moir\'{e} materials, we present \texttt{PARPHOM}, a software package for performing phonon calculations in moir\'{e} systems. \texttt{PARPHOM} utilizes structures relaxed with classical force fields in the \texttt{LAMMPS} \cite{LAMMPS} package to generate the force constant matrix. Subsequently, it employs \texttt{ ScaLAPACK} \cite{slug} routines to diagonalize the dynamical matrix. The utilities provided with \texttt{PARPHOM} can compute the density of states using either the linear triangulation method or by smearing with a Gaussian or Lorentzian function. Moreover \texttt{PARPHOM} can also capture the temperature dependence of the phonon density of states, the shift in the phonon modes non-perturbatively through mode-projected velocity autocorrelations \cite{phonon_quasiparticle,PhysRevB.47.4863}, including the chirality of modes.

\section{Force Constants}
\label{ForceConst}
In a system of $n$ atoms, the total potential energy can be expressed as a function of the atomic positions, $\mathbf{r}_{i}$ as follows:
\begin{equation*}
    V(\mathbf{r}_{1}, \mathbf{r}_{2}, \cdots \mathbf{r}_{n})
\end{equation*}
and the force constants are defined as the second derivative of this potential
\begin{equation}
    \Phi_{\alpha\beta}(jl, j'l') = \frac{\partial^2 V}{\partial r_\alpha(jl)
\partial r_\beta(j'l')} = -\frac{\partial F_\beta(j'l')}{\partial
r_\alpha(jl)}
    \label{force_constant}
\end{equation}
where $j,j'$ are the indices of the atoms in the unit cell and $l,l'$ denote the indices of the unit cells, $\alpha,\beta$ are the Cartesian coordinates and $F_\alpha(jl)$ is the force acting on the $j^{\text{th}}$ particle in the $l^{\text{th}}$ unit cell in the direction $\alpha$.\\
The force constants are generated using a finite displacement method. 
Each atom, is displaced by a finite distance (defined by the user) relative to its equilibrium position obtained from the relaxed \texttt{LAMMPS} structure along a Cartesian direction, and the resultant forces on all the atoms are obtained. 
The derivatives of these forces with respect to the displacements provide the force constants (Eqn. \ref{force_constant}). 
We opt to compute the derivatives through central finite differences. 
To do so, another displacement of the atom in the opposite direction is used to obtain the force constant as:
\begin{equation}
    \Phi_{\alpha\beta}(jl, j'l') \simeq -\frac{ F_\beta(j'l';\Delta r_\alpha{(jl)}/2) - 
    F_\beta(j'l';-\Delta r_\alpha{(jl)}/2)} 
    {\Delta r_\alpha(jl)},
\end{equation}
where $F_\beta(j'l'; \Delta r_\alpha{(jl)})$ are the forces on the $j'$ atom in the $l'$ unit cell due to a finite displacement $\Delta r_\alpha(jl)$ on the $j^{\text{th}}$ atom in the $l^{\text{th}}$ unit cell. \\
It can be shown that the atomic force constants depend on $l$ and $l'$ only thorough their difference \cite{maradudin1968symmetry}
\begin{equation}
    \Phi_{\alpha\beta}(jl, j'l') = \Phi_{\alpha\beta}(j0, j'l'-l)
\end{equation}
Consequently, displacing only the atoms in the $0^{\text{th}}$ cell and computing the forces on the $j'$ atom in all the cells up to a cutoff is sufficient for our purposes.  
A displacement of $\Delta r_\alpha = 10^{-4}$\AA \ in any Cartesian direction is seen to work well in most cases and has been set as the default displacement distance in our code.  

Contemporary classical force fields employ a cutoff length for computing the interatomic forces. Beyond this cutoff length, interatomic forces are considered negligible. This cutoff length typically falls within the range of approximately $10$ \AA. In conventional calculations involving moiré systems, the lattice vectors exceed $10$ nm in length. At these length scales, the cutoff distances of most readily available force fields employed for interatomic force calculations would be exceeded. Hence for almost all moiré systems, it is sufficient to perform the force constant calculations on the moiré unit cell instead of constructing moiré supercells. 

Our method for the generation of force constants via finite differences is embarrassingly parallel since for each atom, $j'$, the elements of the force constant matrix generated by displacing the $j^{\text{th}}$ atom in the direction $\alpha$, can be computed independently. A dataset of dimension $(n,n,3,3)$ is initialized to store our force constant matrix using \verb|h5py| \cite{collette_python_hdf5_2014, h5py}. The generation of force constants is parallelized by distributing the $n$ atoms among the $N_p$ processors over which the calculation is carried out. The displacement for every atom within the moiré cell, in each Cartesian direction, is performed separately to generate the $(n,3)$ dimensional array $\Phi_{\alpha\beta}(j0,j'0)$.  
The data set is then written to a file using parallel-\verb|h5py|.

The currently available codes for phonon calculations, such as \texttt{PHONOPY}\cite{phonopy}, further use symmetry operations to reduce the number of displacements required to generate the force constants. Because our code is massively parallel, the overhead cost to generate the symmetry operators is comparable to the time required to perform all the $6n$ displacements required to generate the complete force constant matrix. As a result, we have chosen not to reduce the number of displacement operations using symmetry but perform the entire set of $6n$ displacements.

Subsequent processing of the generated force constants is necessary to guarantee that the symmetries of the force-constant matrices are satisfied. The first of these symmetries is the transposition or index symmetry, reflecting the invariance under the permutation of indices. This symmetry originates from the properties of the partial derivatives in eqn (\ref{force_constant}):
\begin{equation}
   \Phi_{\alpha\beta}(jl,j'l') = \Phi_{\beta\alpha}(j'l',jl)
\end{equation}
The translational symmetry, which arises from the observation that if all the atoms are translated by an identical amount $(\mathbf{u})$, the energy of the structure remains invariant. Hence the net force acting on the atoms due to such a displacement must be zero in each direction 
\begin{equation}
    \begin{split}
        \sum_{j'l'} F_{\beta}(j'l') &= \sum_{j'l'} -\Phi_{\alpha\beta}(jl,j'l')\Delta r_{\alpha}(jl) 
                       \\ &= - \mathbf{u} \sum_{j'l'} \Phi_{\alpha\beta}(jl,j'l') = 0 \\
        i.e.  \ \ \ \ &\forall (\alpha,\beta,j,l), \ \ \ \ \sum_{j'l'} \Phi_{\alpha\beta}(jl,j'l') = 0 
    \end{split}
\end{equation}
This result is referred to as the acoustic sum rule. Recently, novel methods for implementing these two symmetries simultaneously, like the cluster space algorithm \cite{eriksson2019hiphive}, have been proposed. In our package, we have implemented a computationally inexpensive iterative method, where both symmetries are applied sequentially until the maximum number of iterations is reached.

Additionally, the force constant matrix elements will have the discrete rotational symmetries of the crystal. However, in the moir\'{e} materials which we are interested in, especially in the small angle limit, the relaxed lattice structures do not exhibit the $D_6$ or $D_3$ symmetries present in the rigid structures. This was verified by generating the symmetries via \texttt{Spglib}, which found time-reversal as the only symmetry in the relaxed low angle systems.  Consequently, we have not explicitly enforced the discrete rotational symmetry in the force constants. It is to be noted that when generating the forces via \texttt{LAMMPS}, the discrete rotational symmetries would be encoded up to the some threshold. In most cases, this is sufficient to get the correct rotational symmetries of the force constants without any additional input.

In rare cases where the force field cutoff length exceeds the moiré unit cell length, information about the moiré supercell force constants is necessary to generate an accurate phonon spectrum. Our script's replication feature can generate the force constants in unit cells of materials parameterized with force fields having cutoff distances bigger than the moiré unit cell. The force constants can be generated on a $(m_1 \times m_2 \times m_3)$ moiré supercell using the \verb|-r| flag (only integer values of $m_i$ are permitted). When force constants in replicated structures are requested, the force constant array is stored in a $(N,N,3,3)$ dimensional dataset, where $N$ is the number of atoms in the moiré supercell.

\section{Phonon Dispersion}
\subsection{Dynamical Matrix}
\label{DynMat}
The dynamical matrix of any system at a $\mathbf{q}-$point in the reciprocal space can be computed from the force constants as:
\begin{equation}
    D_{\alpha\beta}(jj',\mathbf{q}) = \frac{1}{\sqrt{m_j m_{j'}}} \sum_{l'}
\Phi_{\alpha\beta}(j0, j'l')
e^{i\mathbf{q}\cdot[\mathbf{r}(j'l')-\mathbf{r}(j0)]}
\end{equation}
When we are computing the force constants within just the moiré unit cell, the dynamical matrix can be written as 
\begin{equation}
    D_{\alpha\beta}(jj',\mathbf{q})  = \frac{1}{\sqrt{m_j m_{j'}}} \Phi_{\alpha\beta}(j0, j'0)f(\mathbf{q})
    \label{eqn:dynmat}
\end{equation}
where f($\mathbf{q}$) is a phase factor given as:
\begin{equation}
    f(\mathbf{q}) = \frac{1}{n_{j'}}\sum_{l} e^{i\mathbf{q}\cdot[\mathbf{r}_{j}(0) - \mathbf{r}_{j'}(l)]}
    \label{eqn:phase}
\end{equation}

The summation in Eqn. (\ref{eqn:phase}) is over all the nearest neighbour cells such that $||\mathbf{r}_{j}(0) - \mathbf{r}_{j'}(l)||$ is minimum, and $n_{j'}$ is the number of image atoms satisfying the minimization criteria across the neighbouring unit cells. This holds true because of the fact that \texttt{LAMMPS} uses the minimum image convention while computing the forces between any pair of atoms, $(j,j')$. Consequently the force constants $\Phi_{\alpha\beta}(j0, j'0)$ would have the sum over the neighbouring unit cells built-in by design, and hence we do not need to sum over the neighbouring unit cells separately. 

Once the dynamical matrix is constructed, it is diagonalized to obtain the phonon eigenvectors $(\Psi_{\mathbf{q}\nu})$ and the eigenvalues $(\omega_{\mathbf{q}\nu})$ at the $\mathbf{q}-$point.
\begin{equation}
    \sum_{j'\beta} D_{\alpha\beta}(jj',\mathbf{q}) \Psi_{\mathbf{q}\nu}^{j'\beta} =
    \omega_{\mathbf{q}\nu}^2 \Psi_{\mathbf{q}\nu}^{j\alpha}
\end{equation}

We use the \texttt{ScaLAPACK} routines to diagonalize this matrix. \texttt{ScaLAPACK} requires the matrix to be distributed in a block-cyclic fashion. From Equations (\ref{eqn:dynmat}) and (\ref{eqn:phase}), it is apparent that every element of the dynamical matrix can be generated locally in each processor independently after reading the appropriate force constant element. Hence, we generate the elements local to each processor for the block-cyclically distributed dynamical matrix in-situ.
Once the dynamical matrix has been generated, we use the routine \verb|PZHEEVX| for diagonalization of Hermitian matrices to obtain the eigenvalues and eigenvectors. 

The user has the option of choosing to compute just the eigenvalues or the eigenvalues along with the eigenvectors. Furthermore, the \verb|PZHEEVX| routine allows the computation of eigenvalues (and the corresponding eigenvectors, if requested) within a certain interval of their values, or the eigenvalues within a range of indices. The appropriate flags for enabling these features are mentioned in the documentation.

\subsection{Group Velocity}
The velocity of a particular phonon mode is
\begin{equation}
\textbf{v}_{\mathbf{q}\nu} = \nabla_{\mathbf{q}}\omega_{\mathbf{q}\nu} = \frac{\partial\omega_{\mathbf{q}\nu}}{\partial \mathbf{q}}   
\end{equation}
One of the most commonly used method for computing the velocities is the one using finite differences. But in this method, the evaluation of the velocity at the band crossings becomes non-trivial as the derivatives are ill-defined. To overcome this issue, we compute the velocities as \cite{togo2015distributions}
\begin{equation}
\begin{split}
    \textbf{v}_{\mathbf{q}\nu} = 
\frac{1}{2\omega_{\mathbf{q}\nu}}\frac{\partial\omega_{\mathbf{q}\nu}^2}{\partial
\mathbf{q}} &=\frac{1}{2\omega_{\mathbf{q}\nu}}\frac{\partial}{\partial\mathbf{q}}\langle\Psi_{\mathbf{q}\nu} \lvert D(\mathbf{q}) \lvert \Psi_{\mathbf{q}\nu} \rangle 
\\ 
&= \frac{1}{2\omega_{\mathbf{q}\nu}} \bigg\langle\Psi_{\mathbf{q}\nu} \bigg\lvert \frac{\partial D(\mathbf{q})}{\partial\mathbf{q}} \bigg\lvert \Psi_{\mathbf{q}\nu} \bigg\rangle
\end{split}
\end{equation}
We provide two avenues for computing the derivative of the dynamical matrix. 
In the first method, the analytic derivative of the dynamical matrix can be calculated from eqn. (\ref{eqn:dynmat}) and (\ref{eqn:phase}) as 
\begin{equation}
    \frac{\partial D_{\alpha\beta}(jj',\mathbf{q})}{\partial \mathbf{q}_{\zeta}} = \frac{i \ \Phi_{\alpha\beta}(j0, j'0)}{n_{j'}\sqrt{m_j m_{j'}}}
		\sum_{l} e^{i\mathbf{q}\cdot[\mathbf{r}_{j}(0) - \mathbf{r}_{j'}(l)]} \big[\mathbf{r}_{j}(0) - \mathbf{r}_{j'}(l)\big]_{\zeta}
\end{equation}
where $\zeta$ is the direction of the derivative and $\big[\mathbf{r}_{j}(0) - \mathbf{r}_{j'}(l)\big]_{\zeta}$ is the component of the vector in the $\zeta-$direction.

In the second method the derivative is evaluated by generating the dynamical matrix at $D(\mathbf{q}\pm\Delta_{\mathbf{q}}/2)$ and taking the central finite difference.  
\begin{equation}
    \frac{\partial D(\mathbf{q})}{\partial \mathbf{q}_{\zeta}} = 
    \frac{D(\mathbf{q}+\Delta_{\mathbf{q}}^{\zeta}/2)-D(\mathbf{q}-\Delta_{\mathbf{q}}^{\zeta}/2)}{\Delta_{\mathbf{q}}^{\zeta}}
\end{equation}
The displacement is taken as $\Delta_{\mathbf{q}}^{\zeta} = 10^{-4}$ \AA${^{-1}}$ in the $\zeta^{\text{th}}$ direction in our code.
Once the derivative is obtained, the velocity matrix elements of the $\nu^{\text{th}}$ band in the direction $\zeta$ are given by
\begin{equation}
    \textbf{v}_{\mathbf{q}\nu}^{\zeta} = \frac{1}{2\omega_{\mathbf{q}\nu}} \Big[\Psi_{\mathbf{q}}^{\dagger} \frac{\partial D(\mathbf{q})}{\partial \mathbf{q}_{\zeta}} \Psi_{\mathbf{q}} \Big]_{(\nu,\nu)^{\text{th}} \text{ element}}
    \label{eqn:vel}
\end{equation}
\section{Installation and Workflow}
\subsection{Installation}
The force constant generation is done through the Python script \verb|get_force_constant.py| and requires the following packages: \verb|numpy|, \verb|h5py|, \verb|mpi4py| and the serial version of the \texttt{LAMMPS} Python wrapper. \\
Building the phonon calculation executable requires linking to the \texttt{MPI}, parallel \texttt{HDF5} and \texttt{ScaLAPACK} libraries. We provide two example makefiles with our code, \verb|Makefile.intel| and \verb|Makefile.aocc| for building the executable with Intel compilers and AMD optimized compilers respectively. 
In addition to these, we also provide a Python package, \verb|pyphutil|, for post processing. It can be installed via the standard method for installing Python packages with either \verb|pip install| or using \verb|setup.py install|. 
\subsection{Workflow}
The first step in using the \texttt{PARPHOM} package is to generate the structures for the systems whose phonon spectra need to be calculated. In this work, we utilized the \texttt{Twister} code \cite{naik2022twister} to generate the rigid twisted structures and subsequently relaxed them using available classical force fields in \texttt{LAMMPS}.
Once the relaxed structures are obtained, the user must prepare a \texttt{LAMMPS} input file for the force constant calculation. This file is used to read the force fields and the structure files through the serial Python wrapper for \texttt{LAMMPS}. Using this \texttt{LAMMPS} input file, the script \verb|get_force_constant.py| generates the force constant matrix, and, if requested, symmetrizes it. \\
The default displacement for the generation of the force constant is set to $10^{-4}$ \AA. It is our observation that for most systems, this displacement generates a force constant matrix which satisfies all the symmetry constraints without any additional processing.
Besides the force constant file and the relaxed \texttt{LAMMPS} structure file, the phonon dispersion calculations require a list of $\mathbf{q}$ points in a format similar to standard codes like \texttt{Quantum \ Espresso} \cite{giannozzi2009quantum} for performing the calculations. The input file has to be prepared by the user in the format given as a sample in the \texttt{README} file.

\subsection{Output File}
\begin{figure}
    \centering
    \includegraphics[width=0.48\textwidth]{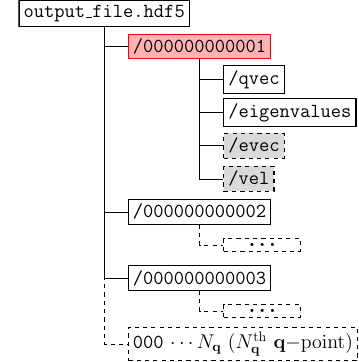}
    \caption{The output file structure for the phonon dispersion. Each output file contains as many groups as there $\mathbf{q}-$points. Every group consists of two mandatory datasets, for the $\mathbf{q}-$vector in crystal coordinates and the eigenvalues. It might also contain two optional datasets, for eigenvectors and group velocity, depending upon the user input}
    \label{fig:OutputFile}
\end{figure}
The output file from the phonon dispersion code is stored in \texttt{HDF5} format in as described below. The file contains a group for each of the $\mathbf{q}-$points for which the calculations are performed. The groups are named as $`/00000000001\text{'}, `/000000000002\text{'},$ and in the same order as the $\mathbf{q}-$points in the input file. If the user has requested $m\le 3n$ eigenvalues and eigenvectors to be calculated, each group would contain the following datasets:

\begin{enumerate}
    \item \verb|/qvec|: $(3,1)$ dimensional array with the $\mathbf{q}-$points in crystal coordinates. 
    \item \verb|/eigenvalues|: $(m,1)$ dimensional array with the phonon frequencies at the $\mathbf{q}-$point in cm$^{-1}$.
    \item \verb|/evec|: Contains the eigenvectors at each $\mathbf{q}-$point. 
    The eigenvector dataset is a $(m\times 3n)$ dimensional array, where $m$ is the number of eigenvectors/eigenvalues requested by the user. These are stored in a compound datatype made up of two double precision data types having their names set as `r' and `i' for the real and imaginary parts respectively. It is to be noted that the each row of the matrix denotes an eigenvector unlike the LAPACK convention of the eigenvectors being stored columnwise.
    \item \verb|/vel|: $(m,3)$ dimensional dataset containing the velocity of each of the bands at the $\mathbf{q}-$point in $m/s$, in each Cartesian direction.
\end{enumerate}

\section{Performance and comparison with existing codes}
\begin{figure}
    \centering
    \includegraphics[width=\linewidth]{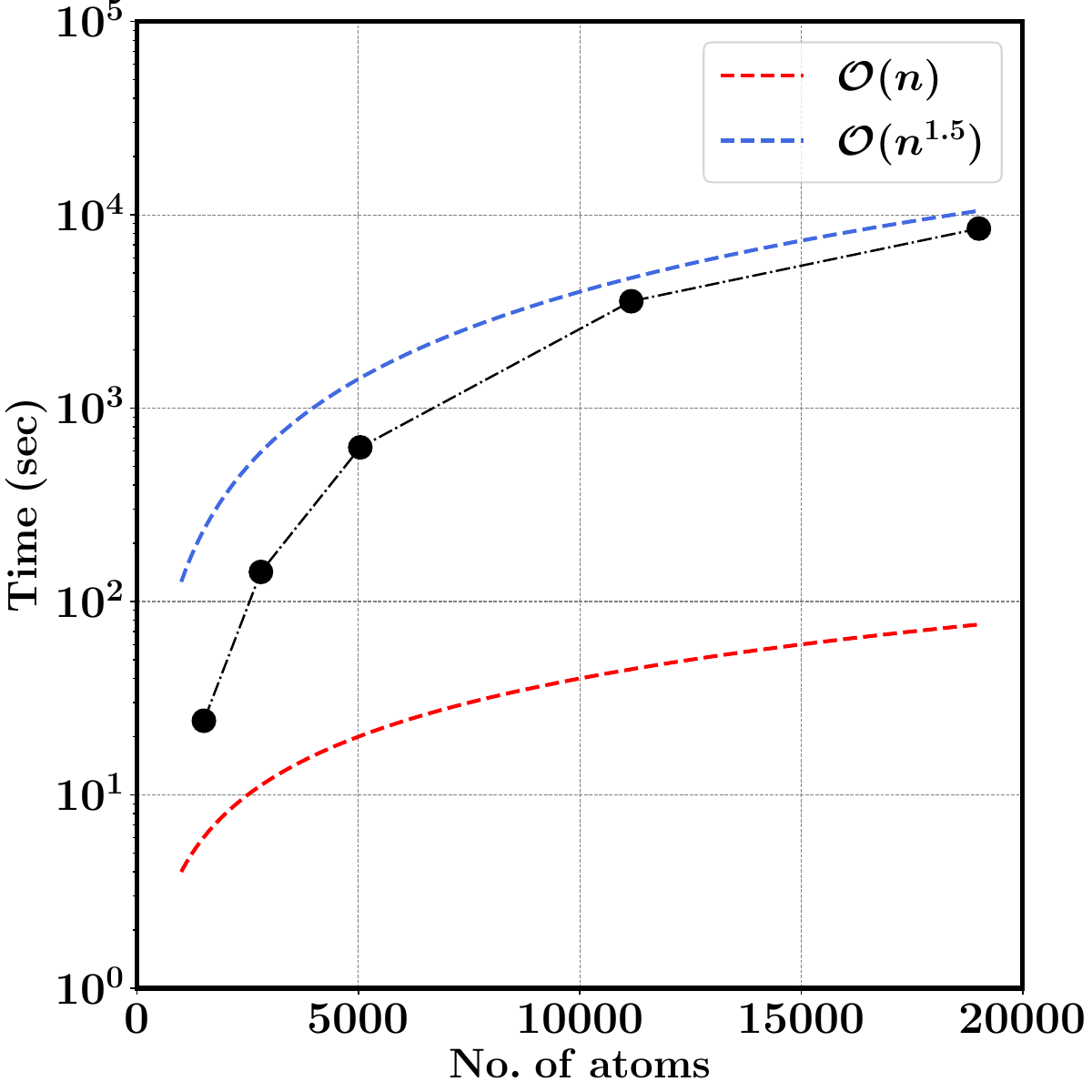}
    \caption{Time taken for generation of Force Constants as a function of atoms in the system. All calculations were performed on $64$ MPI processes.}
    \label{fig:FC_perf}
\end{figure}
One of the major performance bottlenecks in the existing codes 
such as \texttt{PHONOPY} or \texttt{PhonoLAMMPS}, which are routinely used for force constant generation, is that they perform the force constant generation on a single core. We have parallelized the force constant generation to run over multiple processes which naturally gives us significant performance boost in terms of computation time over currently available codes. The time taken to generate the force constant matrices as a function of system size is shown in fig. (\ref{fig:FC_perf}). \\
The benchmarks have been done on a system running on \texttt{Intel Xeon$^{\copyright}$ Gold $6230$} processors with maximum frequency of $3.90$ GHz on each core. All the necessary libraries were built with Intel C compiler, \texttt{icc v$2021.5.0$} and \texttt{Intel MPI v$2021.5.0$} wherever MPI is needed. The \texttt{numpy (v$1.21.2$)}, \texttt{scipy (v$1.6.2$)} and \texttt{LAMMPS (stable\_$29$Sep$2021$\_update$2$)} libraries were linked with \texttt{Intel MKL v$2022.0.1$}. All the benchmark calculations were performed using $64$ MPI processes.

\section{Post Processing}
\begin{figure}
    \centering
    \includegraphics[height=1.2\linewidth]{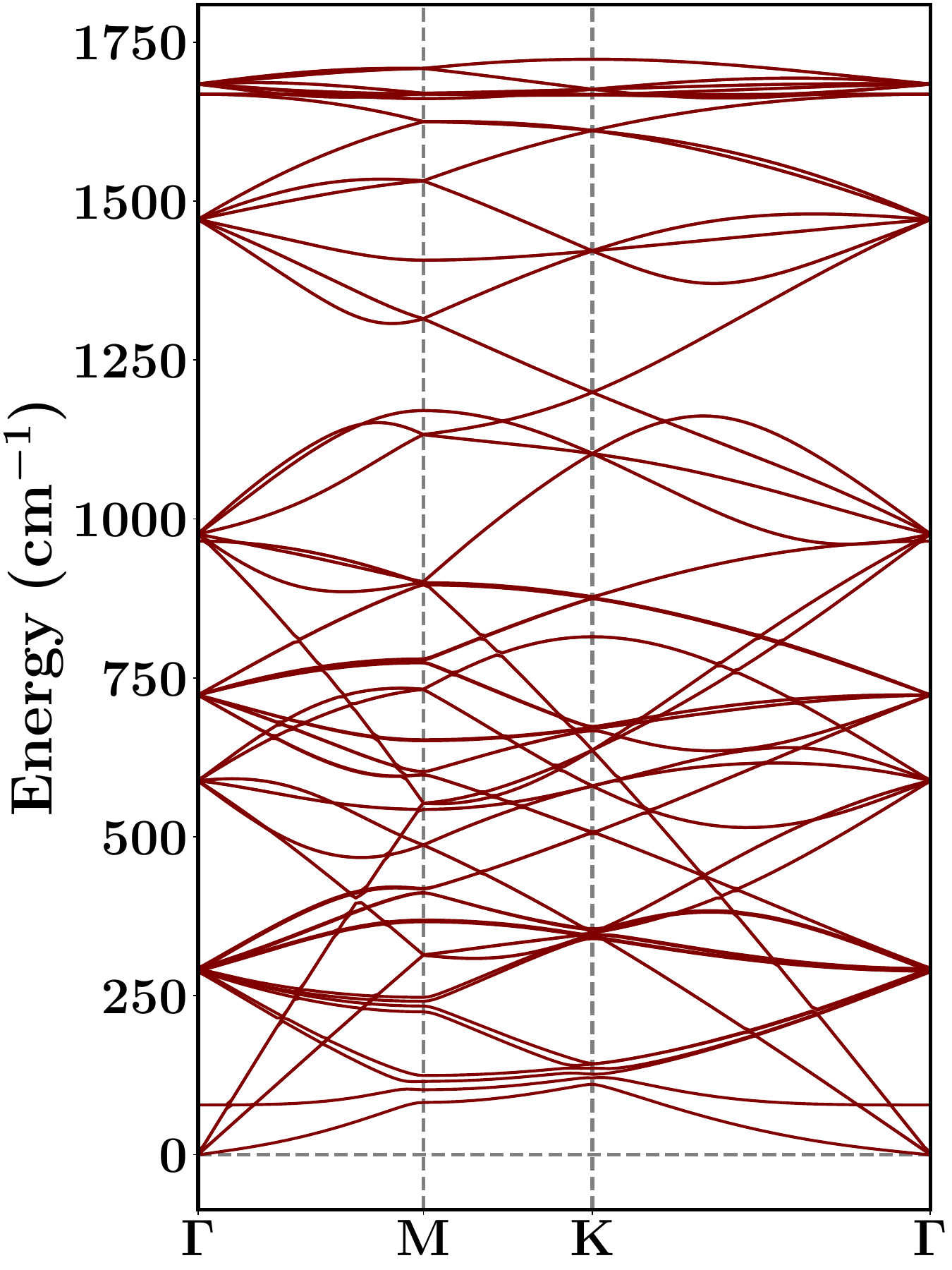}
    \caption{Phonon band structure in $21.8^{\circ}$ twisted bilayer graphene as obtained from our code}
    \label{fig:BS21degtblg}
\end{figure}
Post processing of the data generated in the output file can be performed using the Python package provided with our code, \verb|pyphutil|.
\subsection{Band Structure}
To generate the band structures, the user needs to provide a list of the $\mathbf{q}$ points along the desired path. The \verb|pyphutil| module contains a function to generate $\mathbf{q}$ points along any directions provided by the user. Upon generation of the $\mathbf{q}-$points and the phonon frequencies alone the path, the band structure can be visualized using the function \verb|band_structure| from  the \verb|pyphutil.plot_figures.plot| class. 

The band structure for $21.8^{\circ}$ twisted bilayer graphene obtained along the high symmetry lines as obtained from our code is shown in fig. \ref{fig:BS21degtblg}.
\subsection{Density of States}
\begin{figure}
    \centering
    \includegraphics[width=\linewidth]{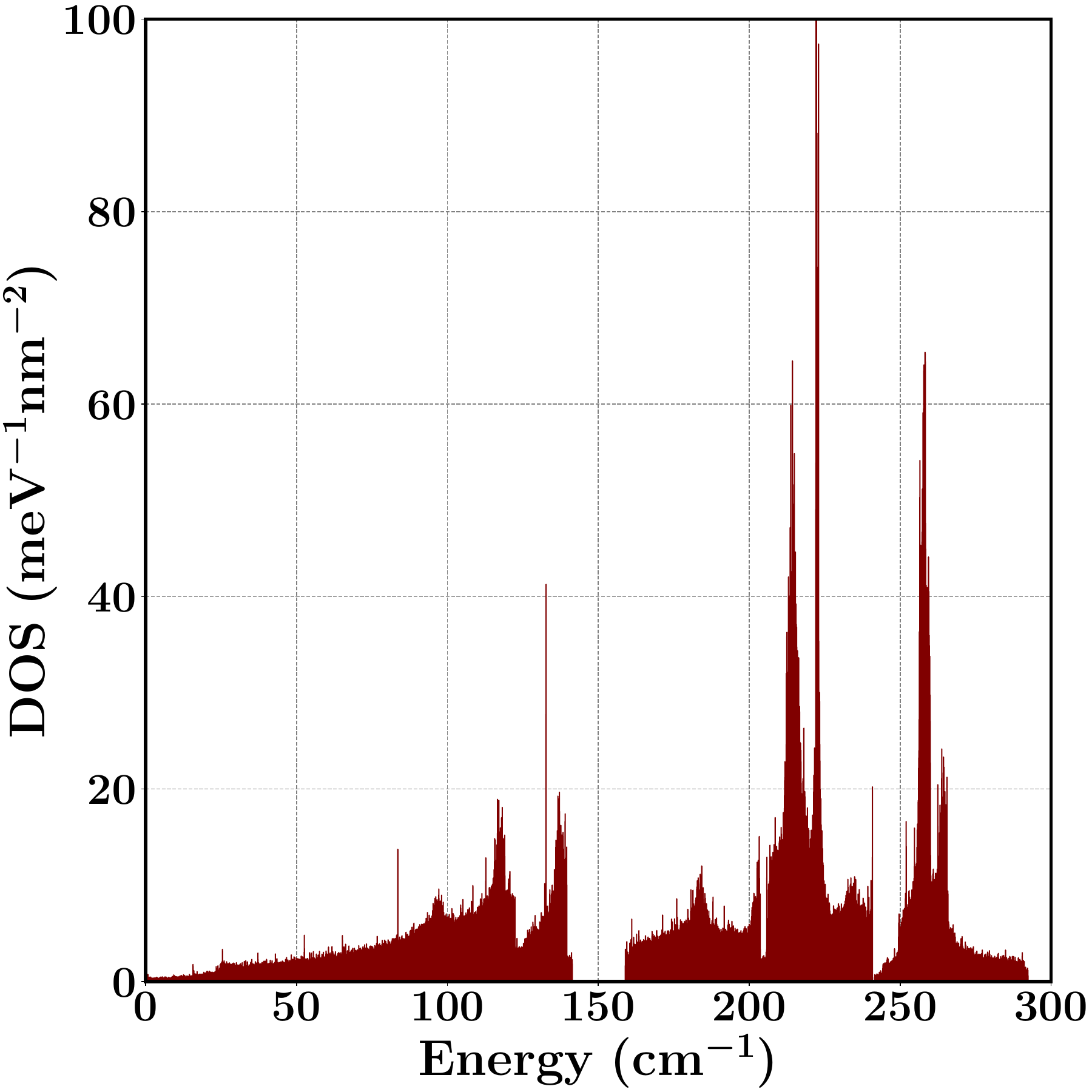}
    \caption{Phonon density of states for $2^{\circ}$ twisted bilayer WSe$_2$ ($4902$ atoms in the moir\'e unit cell), as calculated using our code on a $18\times 18$ $\mathbf{q}-$point grid}
    \label{fig:DOS2degWSe2}
\end{figure}
The density of state computations require the $\mathbf{q}$ points to be available on a uniform grid in the Brillouin zone. Site symmetry operations are typically used to reduce the number of points that need to be sampled. The utility to generate $\mathbf{q}$ points for computing the density of states, uses Spglib \cite{spglib} to find the space group symmetries and construct the reduced $\mathbf{q}$ grid. 
The density of states is computed via the linear triangulation method, which involves breaking up the Brillouin zone into delaunay triangles. This method is described in greater details in \ref{linear triangulation}. The task of evaluating the contribution of each delaunay triangle to total the density of states is independent and hence has been parallelized. The density of states for $2^{\circ}$ twisted bilayer WSe$_{2}$ computed on a $18\times 18$ grid using the linear triangulation method is shown in fig. \ref{fig:DOS2degWSe2}\\
The user also has the option to use Gaussian or Lorentzian smearing instead of the default linear triangulation method for this calculation. For these cases the width of the smearing functions has to be provided by the user.
\\
The density of states routine further allows the user to evaluate Brillouin zone integrals of the forms:
\begin{equation}
    \begin{aligned}
        \text{I}_1(\omega) &= \frac{1}{\Omega_{\text{BZ}}} \sum_{\nu} \int_{\Omega_{\text{BZ}}} f(\mathbf{q},\nu) \Theta(\omega-\omega_{\mathbf{q}\nu}) \ \text{d}\mathbf{q} \\
        \text{I}_2(\omega) &= \frac{1}{\Omega_{\text{BZ}}} \sum_{\nu} \int_{\Omega_{\text{BZ}}} f(\mathbf{q},\nu) \delta(\omega-\omega_{\mathbf{q}\nu}) \ \text{d}\mathbf{q}
    \end{aligned}
\end{equation}
The user can input any $\mathbf{q}-$dependent function and evaluate the corresponding integral over the Brillouin zone.
\section{Additional Utilities}
In addition to the band structure and the Brillouin zone integral evaluations, our package allows the user to look into the finite temperature dynamics of the phonon modes and their chiral nature.
\subsection{Finite temperature dynamics}
The velocity autocorrelation function provides an avenue for calculating the dynamic properties of systems at finite temperatures. This method takes into account all the anharmonic interactions between different atoms as it is based on the processing of the atomic trajectories. \\
The velocity autocorrelation in the $\alpha^{\text{th}}$ direction for the $i^{\text{th}}$ particle with velocity $\mathbf{v}_{i\alpha}(t)$ at time $t$, is defined as
\begin{equation}
    \langle \mathbf{v}_{i\alpha}(0)\mathbf{v}_{i\alpha}(t) \rangle = \lim_{\tau\rightarrow \infty } \frac{1}{\tau}\int_{0}^{\tau} \mathbf{v}_{i\alpha}(t')\mathbf{v}_{i\alpha}(t+t')\text{d}t'
\end{equation}
\subsubsection{Density of states}
\begin{figure}
    \centering
    \includegraphics[width=\linewidth]{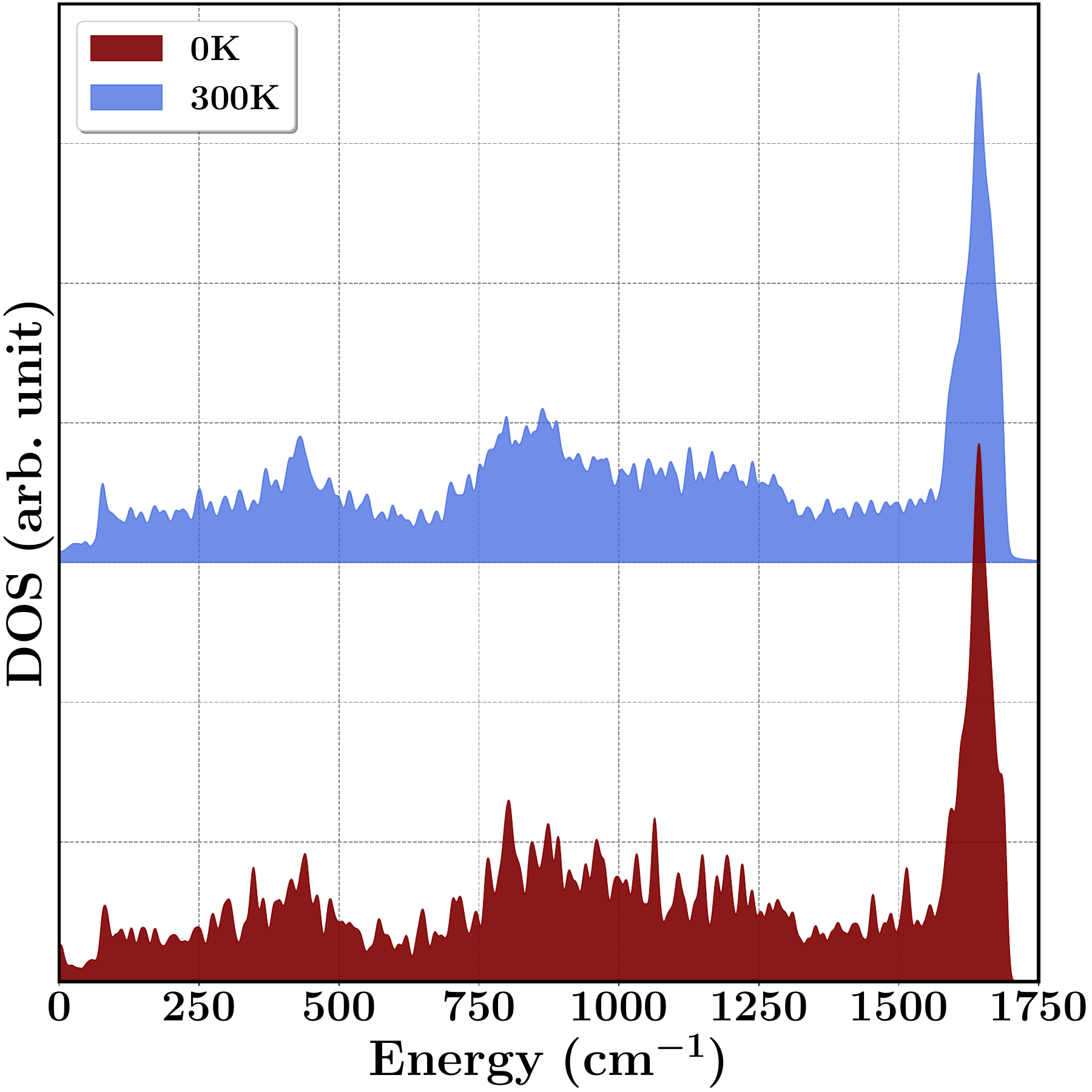}
    \caption{Comparision of density of states for $21.8^{\circ}$ twisted bilayer graphene at $0$K and $300$K.}
    \label{fig:finiteDOS}
\end{figure}
The phonon density of states of the system at a temperature $T$, can be related to the velocity autocorrelation function as \cite{PhysRevB.47.4863} 
\begin{equation}
    \rho(\omega) = \frac{1}{Nk_B T} \sum_{i,\alpha} m_i \int_{0}^{\infty} \langle \mathbf{v}_{i\alpha}(0)\mathbf{v}_{i\alpha}(t) \rangle e^{-i\omega t} dt 
\end{equation}
where $m_i$ is the mass of the $i^{\text{th}}$ particle and $N$ is the number of atoms in the system.
Using the convolution theorem, the expression simplifies to 
\begin{equation}
    \rho(\omega) = \frac{1}{Nk_BT} \sum_{i,\alpha} m_i \lvert \mathbf{v}_{i\alpha}(\omega)  \lvert^2
    \label{eqn:DOS}
\end{equation}
The \verb|finite_temperature| class in the \verb|pyphutil| module is equipped to process the MD trajectory files obtained through LAMMPS for a given temperature and compute the density of states from eqn. (\ref{eqn:DOS}) with the \verb|dos| function that is available. The trajectories have to be generated on a  supercell, large enough in the real space, to ensure a $\mathbf{q}-$point sampling that gives converged results. 

The density of states as obtained from our code for $21.8^{\circ}$ tBLG at $300$K is plotted in fig. \ref{fig:finiteDOS}. The finite temperature trajectories were obtained for a $18\times18$ supercell of the $21.8^{\circ}$ moir\'e unit cell, containing a total of $9072$ atoms. The velocities of all atoms in the production run $(400 \text{ps, in the NVE ensemble \cite{ultrasoft_phason}})$ were collected at every $5$ fs.  
The zero temperature density of states was calculated on a $18\times18$ $\mathbf{q-}$grid.
\subsubsection{Mode Projected velocity autocorrelation}
\begin{figure}
    \centering
    \includegraphics[width=\linewidth]{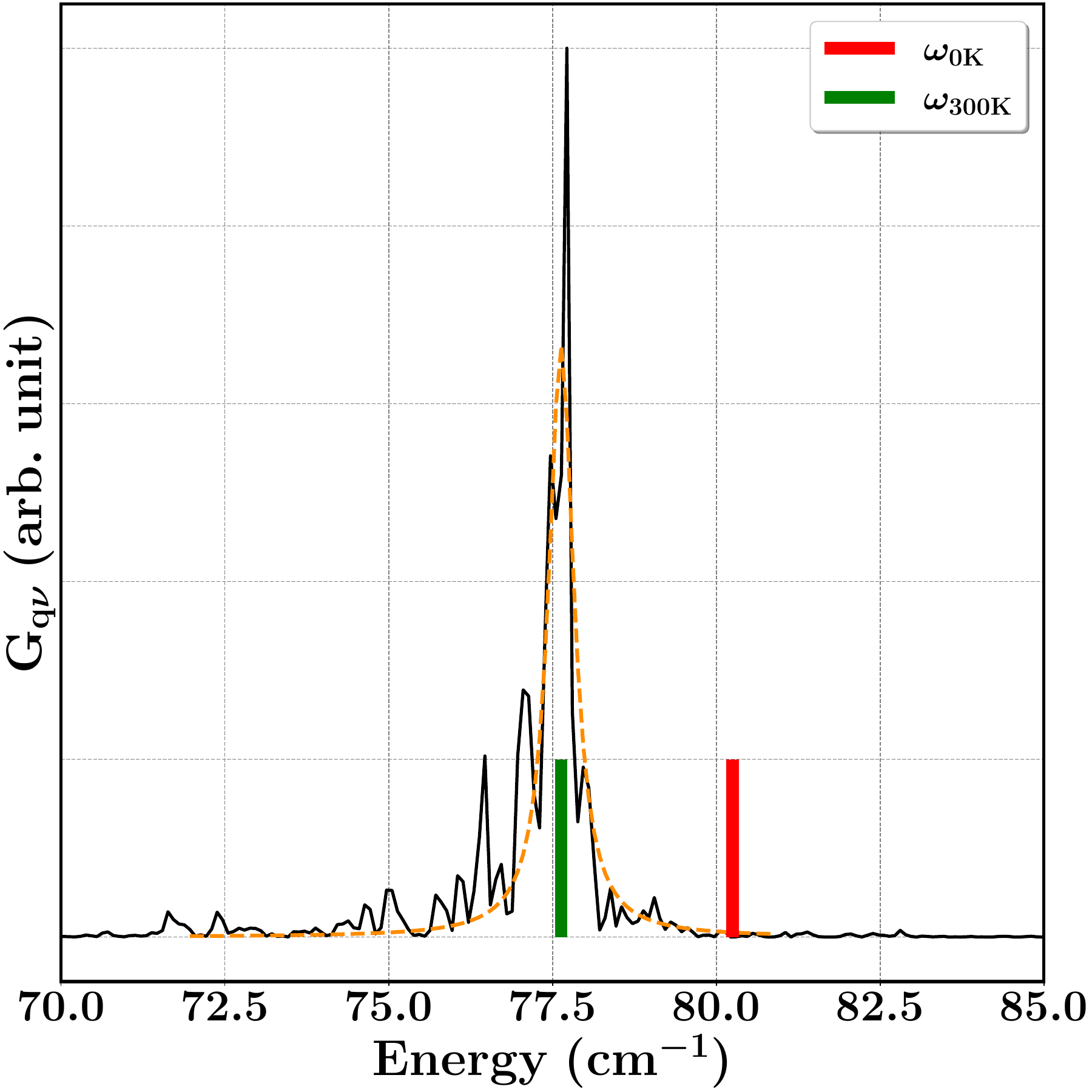}
    \caption{The mode projected power spectrum plotted for the layer breathing mode at the $\Gamma$ point in $1.08^{\circ}$ twisted bilayer graphene at $300$K (in black). The frequency of the mode was obtained by fitting a Lorentzian (in orange) to the spectrum. The center of the fitted Lorentzian is taken as the phonon mode frequency at $300$K (in green). We observe a softening of the mode by $2.6$ cm$^{-1}$ from the $0$K frequency (in red) due to anharmonic effects. }
    \label{fig:Breathing}
\end{figure}
To understand the evolution of individual phonon momdes with temperature, a script to compute the power spectra of mode projected velocity autocorrelation function (MVACF) is also provided as part of the package. 
The MVACF for a particular phonon polarization branch $(\eta_{\mathbf{q}\nu})$ is defined as follows: \cite{phonon_quasiparticle, ultrasoft_phason} 
\begin{equation}
    \langle \text{V}_{\mathbf{q},\nu}(0) \text{V}_{\mathbf{q},\nu}(t) \rangle = \lim_{\tau \to \infty} \frac{1}{\tau} \int_{0}^{\infty}   
    \text{V}_{\mathbf{q},\nu}(t) \text{V}_{\mathbf{q},\nu}^{\mathbf{*}}(t+\tau) \text{d}\tau
\end{equation}
with $\text{V}_{\mathbf{q}\nu}$ denoting the mode projected $\mathbf{q}-$velocities
\begin{equation}
    \begin{aligned}
    \text{V}_{\mathbf{q},\nu}(t) &= \sum_{\mu} \tilde{\mathbf{v}}_{\mathbf{q}}^{\mu}(t)\cdot \eta_{\mathbf{q},\nu}^{\mu} \\
    \tilde{\textbf{v}}_{\mathbf{q}}^{\mu}(t) &= \sqrt{m}_{\mu}\sum_{j} \textbf{v}_{j}(t) \text{e}^{-i \mathbf{q}\cdot\mathbf{r}_{j}(t)}
    \end{aligned}
    \label{eqn:MVACF}
\end{equation}
The index $j$ runs over the atoms of a particular type in the system, while each atom type in the system is denoted by $\mu$. The velocity of the $j^{\text{th}}$ atom of type $\mu$ is denoted as $\mathbf{v}_{j}$. The power spectra of the MVACF 
\begin{equation}
    G_{\mathbf{q}\nu}(\omega) = \int_{0}^{\infty} \langle \text{V}_{\mathbf{q},\nu}(0) \text{V}_{\mathbf{q},\nu}(t) \rangle 
    e^{-i\omega t} dt 
\end{equation}
gives us the information about the linewidths and the lineshifts of the mode, including all the anharmonic effects.

In fig. (\ref{fig:Breathing}) we show the mode projected spectral function of the layer breathing mode at the $\Gamma$ point for $1.08^{\circ}$ twisted bilayer graphene. The trajectories were obtained for a $6\times6$ supercell of the moir\'e unit cell, containing a total of $401904$ atoms. The velocities and positions of all atoms were collected at every $5$ fs for a time period of $400$ps. The power spectra, $G_{\mathbf{q}\nu}(\omega)$ was obtained using the \verb|power_spectra| function in the \verb|finite_temperature| class. The peak of the power spectrum is determined by a Lorentzian fit, shown in orange. The position of the peak denotes the breathing mode frequency at $300$K. The layer breathing mode frequency of bilayer graphene at $0$K is determined to be $80.2$ cm$^{-1}$ and at $300$K we determine the mode frequency from fit as $77.6$ cm$^{-1}$ showing a mode softening of $\sim2.6$ cm$^{-1}$.

\subsection{Chirality of the Phonon Modes}
The existence of chiral phonon modes at the $K$ and $K'$ valleys in TMDs has been predicted \cite{Zhang-Niu} and experimentally verified \cite{chen2015helicity}. Their presence in moir\'e materials has also been shown recently \cite{suri2021chiral,maity2022chiral}. We provide an utility to compute the chirality of the moir\'e phonon modes as part of our package. 

\begin{figure}
    \centering
    \includegraphics[width=\linewidth]{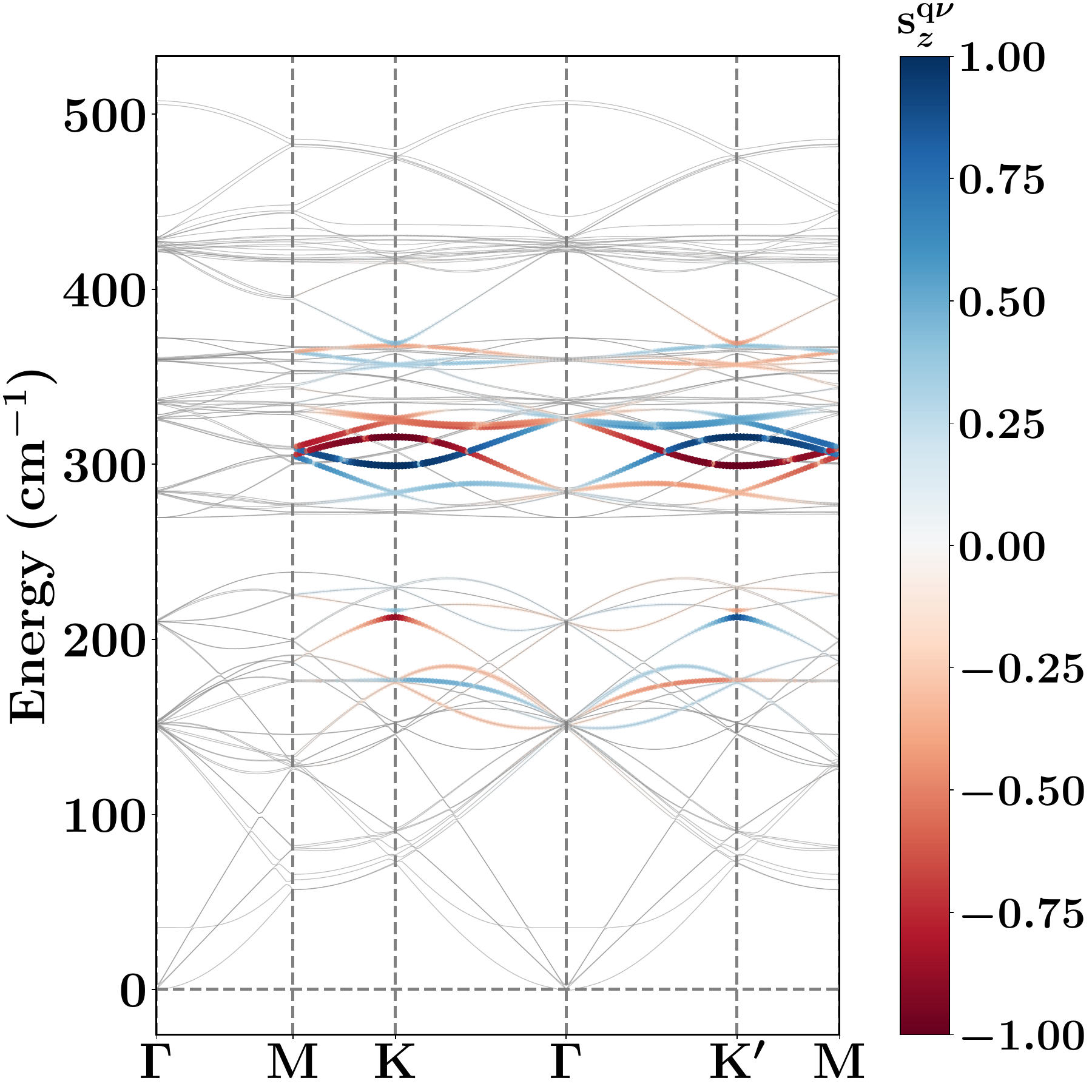}
    \caption{Chirality of the phonon modes in $38.2^{\circ}$ twisted bilayer MoS$_{2}$ about the $z$ direction. The left chiral modes are denoted in blue while the right chiral modes are marked in red.}
    \label{fig:Chirality}
\end{figure}
For computing the chirality along the direction $\alpha$, we consider the basis vectors that decompose a vector into it's right polarized and left polarized components in the plane perpendicular to the $\alpha^{\text{th}}$ direction. As an example, to compute the chirality in $\mathbf{z}-$direction, the basis vectors are the $3n$ dimensional vectors:
\begin{equation}
    \lvert \mathbf{R} \rangle_{\mathbf{z}} = \frac{1}{\sqrt{2n}}
    \begin{bmatrix}
    1 \\
    i \\
    0 
    \end{bmatrix}_{(n)}; \ \ \ \ \ 
    \lvert \mathbf{L} \rangle_{\mathbf{z}} = \frac{1}{\sqrt{2n}}
    \begin{bmatrix}
    1 \\
    -i \\
    0
    \end{bmatrix}_{(n)}
\end{equation}
where $[C]_{(n)}$ denotes the column block $C$ being repeated $n$ times. It is to be noted that these two vectors form a complete basis in the plane perpendicular to the direction of the chirality. \\
For a phonon mode with polarization $\eta_{\mathbf{q}\nu}$, the chirality in the $\alpha^{\text{th}}$ direction is defined as 
\begin{equation}
    \mathbf{s}_{\alpha}^{\mathbf{q}\nu} = \langle \eta_{\mathbf{q}\nu} \lvert  \hat{\mathbf{S}}_{\alpha} \lvert \eta_{\mathbf{q}\nu}  \rangle ; \ \ \ \hat{\mathbf{S}}_{\alpha} = \lvert \mathbf{R} \rangle_{\alpha} \langle \mathbf{R} \lvert_{\alpha} -
    \lvert \mathbf{L} \rangle_{\alpha} \langle \mathbf{L} \lvert_{\alpha}
\end{equation}
In our formulation this is equivalent to evaluating the following expression as shown in \ref{chirality_appendix}:
\begin{equation}
        \mathbf{s}_{\alpha}^{\mathbf{q}\nu} = 2\big[ \mathfrak{Re}(\eta_{\mathbf{q}\nu}^{\beta})\mathfrak{Im}(\eta_{\mathbf{q}\nu}^{\gamma}) - \mathfrak{Im}(\eta_{\mathbf{q}\nu}^{\beta})\mathfrak{Re}(\eta_{\mathbf{q}\nu}^{\gamma})\big]
        \label{eqn:chirality}
\end{equation}
Here $(\alpha,\beta,\gamma)$ denote cyclic permutations of the cartesian coordinates $(x,y,z)$; $\eta_{\mathbf{q}\nu}^{\alpha}$ is the $\alpha^{\text{th}}$ component of the polarization vector $\eta_{\mathbf{q}\nu}$ and $\mathfrak{Re}(\mathbf{a})$ and $\mathfrak{Im}(\mathbf{a})$ denote the real and imaginary part of $\mathbf{a}$.
The phonon mode is linearly polarized in the $\alpha^{\text{th}}$ direction when $\mathbf{s}_{\alpha}^{\mathbf{q}\nu} = 0$, circularly polarized when $\mathbf{s}_{\alpha}^{\mathbf{q}\nu} = 1$ and elliptically polarized when $0 < \lvert \mathbf{s}_{\alpha}^{\mathbf{q}\nu} \rvert < 1 $. \\
In our utility module, we provide the function \verb|chirality| to compute the chirality of the moir\'e modes in all the three Cartesian directions. 
A plot of the chirality of the modes about the $z$ direction in $38.2^{\circ}$ twisted bilayer MoS$_{2}$, evaluated using this module, is shown in fig. \ref{fig:Chirality}. We observe the appearance of chiral phonon modes near the $\mathbf{K}$ and $\mathbf{K'}$ valleys as predicted \cite{maity2022chiral,suri2021chiral}.

\section{Summary}
In this article we present a massively parallel package which can generate the force constants of $2$D systems using classical force fields. A package to calculate the  phonon spectra from these force constants is also being made available. This allows the user to perform phonon calculations on systems of sizes much greater than $10,000$ atoms, in a reasonable amount of time, which is not possible with any currently available package. Furthermore, the utilities provided along with the package enable the user to compute the velocity autocorrelation function, providing information about the finite temperature characteristics of the phonon modes of the systems. Our package also allows the user to compute the chiral nature of each of the phonon mode in $2$D systems. 

\section{Acknowledgements}
M.J. acknowledges the National Supercomputing Mission of the Department of Science and Technology, India, and Nano Mission of the Department of Science and Technology for financial support under Grants No. DST/NSM/R\&D\_HPC\_Applications/2021/23 and No. DST/NM/TUE/QM-10/2019 respectively. H.R.K. acknowledges the Science and Engineering Research Board of the Department of Science and Technology, India, and the Indian National Science Academy for support under Grants No. SB/DF/005/2017 and No. INSA/SP/SS/2023/ respectively. 
H.R.K. also acknowledges support in ICTS by a grant from the Simons Foundation (677895, R.G.).
I.M. acknowledges funding from the European Union's Horizon 2020 research and innovation program under the Marie Skłodowska-Curie Grant agreement No. 101028468.


\appendix

\section{Linear Triangulation Method}
\label{linear triangulation}
 The density of states are calculated using the linear triangular method, a derivative of the linear tetrahedron method in 3D systems.
 
 Suppose we have to evaluate the following integral in the BZ:
	 \begin{equation}
	 	\mathbf{I}(\text{E}) = \frac{1}{\Omega_{BZ}}\sum_{n}\int_{\Omega_{BZ}}  d\mathbf{k} \ f(\mathbf{k}) \Theta(\text{E}-\epsilon_{n\mathbf{k}})
	 	\label{Appeq1}
	 \end{equation}
 and we have the energies, $\epsilon_{n\mathbf{k}}$ at a finite set of $\mathbf{k}$ points, $\{ \mathbf{k}_1, \mathbf{k}_2, \cdots \mathbf{k}_N \}$.
Keeping in tune with the tetrahedron method, we construct delaunay triangles among the set of $\mathbf{k}$ points, and try to evaluate the integral in each of the $N_T$ triangles that are formed. Rewriting eqn.(\ref{Appeq1}) we get,
 	 \begin{equation}
 	 	\begin{aligned}
 	 	\mathbf{I}(\text{E}) &= \frac{1}{\Omega_{BZ}}\sum_{n}\int_{\Omega_{BZ}}  d\mathbf{k} \ f(\mathbf{k}) \Theta(\text{E}-\epsilon_{n\mathbf{k}})\\
 	 	&= \sum_{N_T}\sum_{n}\Big(\frac{1}{\Omega_{BZ}}\int_{\Omega_{T_i}}  d\mathbf{k} \ f(\mathbf{k}) \Theta(\text{E}-\epsilon_{n\mathbf{k}})\Big)\\
 	 	&= \sum_{N_T}\sum_{n} \mathbf{I}_{nT_i}(\text{E})
 	 	\end{aligned}
 	 	\label{Appeq2}
 	 \end{equation}
where $T_i$ denotes the $i^{th}$ triangle.
Suppose we consider a triangle formed with the points $\mathbf{k}_0$, $\mathbf{k}_1$ and $\mathbf{k}_2$, with energies in the $n^{th}$ band as $\epsilon_{n0}$, $ \epsilon_{n1}$ and $\epsilon_{n2}$ respectively. 
Without loss of generality we can assume that the energies are ordered as $\epsilon_{n0} < \epsilon_{n1} < \epsilon_{n2}$. Then in the following cases, the values of the integrals are:
\begin{itemize}[align=left]
    \item[\textbf{(Case 1)} $\text{E} < \epsilon_{n0} < \epsilon_{n1} < \epsilon_{n2}$]\begin{equation}
        \mathbf{I}_{nT_i} = 0
    \end{equation}
    \item[\textbf{(Case 2)} $\epsilon_{n0} \le \text{E} < \epsilon_{n1} < \epsilon_{n2}$] 
    \begin{equation}
      		\begin{aligned}
      			\mathbf{I}_{nT_i}(\text{E}) = \frac{\Delta_{10}\Delta_{20}}{3N_{T_i}}
      			\Big[ (&3-\Delta_{10}-\Delta_{20})f(\mathbf{k_0}) \\&+ \Delta_{10}f(\mathbf{k_{1}})+
      			 \Delta_{20}f(\mathbf{k_{2}})\Big]
      		\end{aligned}
    \end{equation}where $$\Delta_{10} = \Big(\frac{\text{E}-\epsilon_{n0}}{\epsilon_{n1} - \epsilon_{n0}}\Big); \ \ \ \ \Delta_{20} = \Big(\frac{\text{E}-\epsilon_{n0}}{\epsilon_{n2} - \epsilon_{n0}}\Big)$$
    \item[\textbf{(Case 3)} $\epsilon_{n0}  < \epsilon_{n1} \le \text{E} < \epsilon_{n2}$]
    \begin{equation}
      		\begin{aligned}
      		 	\mathbf{I}_{T_i}(\text{E}) = &  \frac{1}{3N_{T_i}}\Big[\big(1-\Delta_{12}\Delta_{02}^2\big)f(\mathbf{k_{0}})+ \big(1-\Delta_{12}^2\Delta_{02}\big)f(\mathbf{k_1}) \\&+\big( 1-3\Delta_{12}\Delta_{02}+ \Delta_{12}\Delta_{02}^2+\Delta_{12}^2\Delta_{02}\big)f(\mathbf{k_2}) \Big]
      		\end{aligned}
      \end{equation}where $$\Delta_{02} = \Big(\frac{\epsilon_{n2}-\text{E}}{\epsilon_{n2} - \epsilon_{n0}}\Big); \ \ \ \ \Delta_{12} = \Big(\frac{\epsilon_{n2}-\text{E}}{\epsilon_{n2} - \epsilon_{n1}}\Big)$$
    \item[\textbf{(Case 4)} $\epsilon_{n0} < \epsilon_{n1} < \epsilon_{n2} \le \text{E}$]
    \begin{equation}
      	\mathbf{I}_{nT_i}(\text{E})	= \frac{1}{3N_{T}} \Big[ f(\mathbf{k_0}) + f(\mathbf{k_1})+f(\mathbf{k_2})\Big]
  	\end{equation}
\end{itemize}

If we are to evaluate inetgrals of the type 
\begin{equation}
    \begin{aligned}
 	    \bm{\rho}(\text{E}) &= \frac{1}{\Omega_{BZ}}\sum_{n}\int_{\Omega_{BZ}}  d\mathbf{k} \ f(\mathbf{k}) \delta(\text{E}-\epsilon_{n\mathbf{k}}) \\ &= \frac{d\mathbf{I}(\text{E})}{d\text{E}} = \sum_{N_T}\sum_{n}\frac{d\mathbf{I}_{nT_i}(\text{E})}{d\text{E}} \equiv \sum_{N_T}\sum_{n}\bm{\rho}_{nT_i}(\text{E}) 	
    \end{aligned}
\end{equation}
we follow the same procedure as before and simply take the derivatves of $\mathbf{I}_{nT_i}(\text{E})$ with respect to E for each case. Hence in 
\begin{itemize}[align=left]
    \item[\textbf{(Case 1)} $\text{E} < \epsilon_{n0} < \epsilon_{n1} < \epsilon_{n2}$]
    \begin{equation}
        \bm{\rho}_{nT_i}(\text{E}) = 0
    \end{equation}
    \item[\textbf{(Case 2)} $\epsilon_{n0} \le \text{E} < \epsilon_{n1} < \epsilon_{n2}$] 
    \begin{equation}
      		\begin{aligned}
      			\bm{\rho}_{nT_i}(\text{E}) =&\frac{1}{3N_{T_i}} 
      		\Big[\Delta_{10}^{\prime}\Delta_{20}\big[(3-\Delta_{10}-\Delta_{20})f(\mathbf{k_{0}}) \\ &+\Delta_{10}f(\mathbf{k_{1}})+\Delta_{20}f(\mathbf{k_{2}})\big] +\Delta_{10}\Delta_{20}^{\prime}[(3-\Delta_{10} \\&-\Delta_{20})f(\mathbf{k_{0}})+\Delta_{10}f(\mathbf{k_{1}})
      		+\Delta_{20}f(\mathbf{k_{2}})] \\&+  \Delta_{10}\Delta_{20}[(-\Delta_{10}^{\prime}-\Delta_{20}^{\prime})f(\mathbf{k_{0}})+\Delta_{10}^{\prime}f(\mathbf{k_{1}})\\&+\Delta_{20}^{\prime}f(\mathbf{k_{2}})]
      		\Big]
      		\end{aligned}
    \end{equation}
    \item[\textbf{(Case 3)} $\epsilon_{n0}  < \epsilon_{n1} \le \text{E} < \epsilon_{n2}$]
    \begin{equation}
      		\begin{aligned}
      		 	\bm{\rho}_{nT_i}(\text{E}) =&\frac{1}{3N_{T_i}}
      	\Bigg[
      	  \big[-2\Delta_{02}\Delta_{02}^{\prime}\Delta_{12}-\Delta_{02}^2\Delta_{12}^{\prime}\big]f(\mathbf{k_{0}})+ \\& \big[-\Delta_{02}^{\prime}\Delta_{12}^2-2\Delta_{02}\Delta_{12}^{\prime}\Delta_{12}\big]f(\mathbf{k_{1}})+ \\&
      	  \big[-3\Delta_{02}^{\prime}\Delta_{12}-3\Delta_{02}\Delta_{12}^{\prime}+2\Delta_{02}\Delta_{02}^{\prime}\Delta_{12}+\\&
      	  \Delta_{02}^2\Delta_{12}^{\prime}+2\Delta_{02}\Delta_{12}^{\prime}\Delta_{12}+\Delta_{02}^{\prime}\Delta_{12}^2\big]f(\mathbf{k_{2}})  \Bigg]
      		\end{aligned}
      \end{equation}
    \item[\textbf{(Case 4)} $\epsilon_{n0} < \epsilon_{n1} < \epsilon_{n2} \le \text{E}$]
    \begin{equation}
      	\bm{\rho}_{nT_i}(\text{E}) = 0
  	\end{equation}
\end{itemize}

where $\Delta_{ij}^{\prime} = \frac{d\Delta_{ij}}{d\text{E}}$ and the $\Delta_{ij}$'s are defined as before. 

\section{Expressions for chirality of modes}
\label{chirality_appendix}
We have taken the right circular and the left circular basis for the $\mathbf{z}-$polarization as
\begin{equation}
    \begin{aligned}
        \lvert \mathbf{R} \rangle_{\mathbf{z}} = \frac{1}{\sqrt{2n}}
        \begin{bmatrix}
        1 \\
        i \\
        0 
        \end{bmatrix}_{(n)}; \ \ \ \ \ 
        \lvert \mathbf{L} \rangle_{\mathbf{z}} = \frac{1}{\sqrt{2n}}
        \begin{bmatrix}
        1 \\
        -i \\
        0
    \end{bmatrix}_{(n)}
    \end{aligned}
\end{equation}
with $[C]_{(n)}$ denoting the column block $C$ being repeated $n$ times as defined in the main text. \\
Then, 
\begin{equation}
    \begin{aligned}
    \lvert \mathbf{R} \rangle_{\mathbf{z}} \langle \mathbf{R} \lvert_{\mathbf{z}} &= \frac{1}{2n} \ \mathbf{I}_{n} \otimes \begin{bmatrix}
    1 && -i && 0 \\
    i && 1 && 0 \\
    0 && 0 && 0
    \end{bmatrix} \\
    \lvert \mathbf{L} \rangle_{\mathbf{z}} \langle \mathbf{L} \lvert_{\mathbf{z}} &= \frac{1}{2n} \  \mathbf{I}_{n} \otimes \begin{bmatrix}
    1 && i && 0 \\
    -i && 1 && 0 \\
    0 && 0 && 0
    \end{bmatrix}   
    \end{aligned}
\end{equation}
Correspondingly, the expression for the chirality operator in the $\mathbf{z}-$direction is
\begin{equation}
    \hat{\mathbf{S}}_{\mathbf{z}} =  \mathbf{I}_{n} \otimes 
    \begin{bmatrix}
    0 && -i && 0 \\
    i && 0  && 0 \\
    0 && 0 && 0
    \end{bmatrix}
\end{equation}
If we denote the mode $\eta_{\mathbf{q}\nu} = [\eta_{\mathbf{q}\nu}^{1x},  \eta_{\mathbf{q}\nu}^{1y}, \eta_{\mathbf{q}\nu}^{1z}, \cdots, \eta_{\mathbf{q}\nu}^{nx},  \eta_{\mathbf{q}\nu}^{ny}, \eta_{\mathbf{q}\nu}^{nz} ]^T$,
the $\mathbf{z}-$chirality can be expressed as
\begin{equation}
    \begin{aligned}
        \mathbf{s}_{\mathbf{z}}^{\mathbf{q}\nu} &= \langle \eta_{\mathbf{q}\nu} \lvert  \hat{\mathbf{S}}_{\alpha} \lvert \eta_{\mathbf{q}\nu}  \rangle \\
        &= \sum_{p=1}^{n} -i (\eta_{\mathbf{q}\nu}^{px})^{*}\eta_{\mathbf{q}\nu}^{py} + i (\eta_{\mathbf{q}\nu}^{py})^{*}\eta_{\mathbf{q}\nu}^{px}
    \end{aligned}
\end{equation}
If we denote the mode polarization $\eta_{\mathbf{q}\nu}^{\alpha} = [\eta_{\mathbf{q}\nu}^{1\alpha},  \eta_{\mathbf{q}\nu}^{2\alpha}, \cdots, \eta_{\mathbf{q}\nu}^{n\alpha}]^T$ as the $\alpha$ component of the mode, it is easy to see that 
\begin{equation}
        \mathbf{s}_{\mathbf{z}}^{\mathbf{q}\nu} = 2\big[ \mathfrak{Re}(\eta_{\mathbf{q}\nu}^{\mathbf{x}})\mathfrak{Im}(\eta_{\mathbf{q}\nu}^{\mathbf{y}}) - \mathfrak{Im}(\eta_{\mathbf{q}\nu}^{\mathbf{x}})\mathfrak{Re}(\eta_{\mathbf{q}\nu}^{\mathbf{y}})\big]
\end{equation}
Similar analysis for the other two directions, with the right and left circular basis vectors modified appropriately, give
\begin{equation}
    \begin{aligned}
        \mathbf{s}_{\mathbf{y}}^{\mathbf{q}\nu} &= 2\big[ \mathfrak{Re}(\eta_{\mathbf{q}\nu}^{\mathbf{z}})\mathfrak{Im}(\eta_{\mathbf{q}\nu}^{\mathbf{x}}) - \mathfrak{Im}(\eta_{\mathbf{q}\nu}^{\mathbf{z}})\mathfrak{Re}(\eta_{\mathbf{q}\nu}^{\mathbf{x}})\big] \\
        \mathbf{s}_{\mathbf{x}}^{\mathbf{q}\nu} &= 2\big[ \mathfrak{Re}(\eta_{\mathbf{q}\nu}^{\mathbf{y}})\mathfrak{Im}(\eta_{\mathbf{q}\nu}^{\mathbf{z}}) - \mathfrak{Im}(\eta_{\mathbf{q}\nu}^{\mathbf{y}})\mathfrak{Re}(\eta_{\mathbf{q}\nu}^{\mathbf{z}})\big]
    \end{aligned}
\end{equation}

\section{Details of calculations}
For the calculations on twisted bilayer graphene, the system was relaxed in \texttt{LAMMPS} using DRIP \cite{wen2018dihedral} as the interlayer potential and REBO \cite{brenner2002second} as the intralayer potential for fig. (\ref{fig:BS21degtblg}) and fig(\ref{fig:Breathing}), and Tersoff \cite{lindsay2010optimized} for the intralayer potential in fig. (\ref{fig:finiteDOS}). For the calculations on twisted TMDs the in plane potential was simulated using the Stillinger Weber potential \cite{mobaraki2018validation, kandemir2016thermal} and the interlayer interaction is modelled using a Kolmogorov Crespi potential \cite{naik2019kolmogorov}. The systems were relaxed by minimization of forces upto a threshold of $10^{-6}$ eV/\AA. The force constants for all the systems were generated with a displacement of $10^{-4}$\AA.


\end{document}